\documentclass[12pt]{iopart}
\usepackage{graphicx}
\begin{document}

\title[Electromagnetically induced transparency and
absorption]{Electromagnetically induced transparency and
absorption cross-over with a four-level Rydberg system}

\author{Ya\u{g}{\i}z Oyun$^1$, \"{O}zg\"{u}r \c{C}ak{\i}r$^2$, Sevilay Sevin\c{c}li$^{1,*}$}

\address{$^1$Department of Photonics, \.{I}zmir Institute of Technology, G\"{u}lbah\c{c}e Campus, Urla,
\.{I}zmir, Turkey}
\address{$^2$Department of Physics, \.{I}zmir Institute of Technology, G\"{u}lbah\c{c}e Campus, Urla,
\.{I}zmir, Turkey}

\ead{sevilaysevincli@iyte.edu.tr}
\vspace{10pt}
\begin{indented}
\item[]
\end{indented}

\begin{abstract}

Electromagnetically induced transparency (EIT) and absorption (EIA) are quantum coherence phenomena which result from the interference of excitation pathways
and combining these with Rydberg atoms have opened up many possibilities for various applications. We introduce a theoretical model to study Rydberg-EIT and Rydberg-EIA effects in cold Cs and Rb atomic ensembles in a four-level ladder type scheme taking into account van der Waals type interactions between the atoms. Proposed many-body method for analysis of such systems involves a self-consistent mean field approach and, it produces results which display a very good agreement with recent experiments. Our calculations also successfully demonstrate experimentally observed EIT-EIA cross-over in Rb case. Being able to simulate the interaction effects in such systems has significant importance, especially for controlling the optical response of these.

\end{abstract}

%
\vspace{2pc}
\noindent{\it Keywords}: Rydberg atoms, Electromagnetically Induced Transparency, Electromagnetically Induced Absorption, Three-photon excitation

%
\maketitle
%
%

\section{Introduction}
More than two decades after the experimental availability of cold gases, these systems yielded many important results and have been used in the fields from quantum optics to quantum information and from nonlinear optics to quantum simulation \cite{bloch05,bloch12}. Combining cold atom systems with electromagnetically induced transparency (EIT) phenomenon that makes an opaque medium transparent under certain conditions via dark-state formation \cite{harris90} resulted in many exciting findings such as slow light \cite{hau99} and light storage in a medium \cite{fleis00,fleis05}. Being a complementary effect to EIT, electromagnetically induced absorption (EIA) is the increase in absorption via atomic coherences and it is referred to as the bright resonance\cite{akul98, lez99}. It has been proposed that EIA can be utilizable for many applications such as optical switches, information storing and magnetometry \cite{ lez99,laz15}. Cold Rydberg gases, consisting of highly excited atoms have also attracted much attention in the last two decades \cite{bloch12,saff10}. Extraordinary properties of Rydberg atoms \cite{gallagher}, including strong interactions with each other and external fields and, Rydberg blockade effect \cite{gaet09,urb09} have provided a good platform for quantum information applications \cite{saff10, comp10}. It has been shown by many experimental and theoretical studies that they are also good candidates for nonlinear optics applications \cite{pritch13,stan13}. Rydberg-EIT systems have provided a significant increase in third-order susceptibility \cite{pritch13,moha08,sev11b} and therefore can be used to achieve strong photon-photon interactions. Combining quantum coherence effects (EIT and EIA) with extraordinary properties of Rydberg atoms, makes it possible to obtain higher order nonlinearities and gives rise to many new opportunities \cite{gart13,ates11,bhat15,gaul16,han15,peyron12,sev11}.  It has been expected that these properties would be very important for single photon sources and all optical switches and as a result they might yield more exciting effects \cite{baur14}.

\begin{figure}[!htbp]
    \centering
	\includegraphics[width= 8 cm]{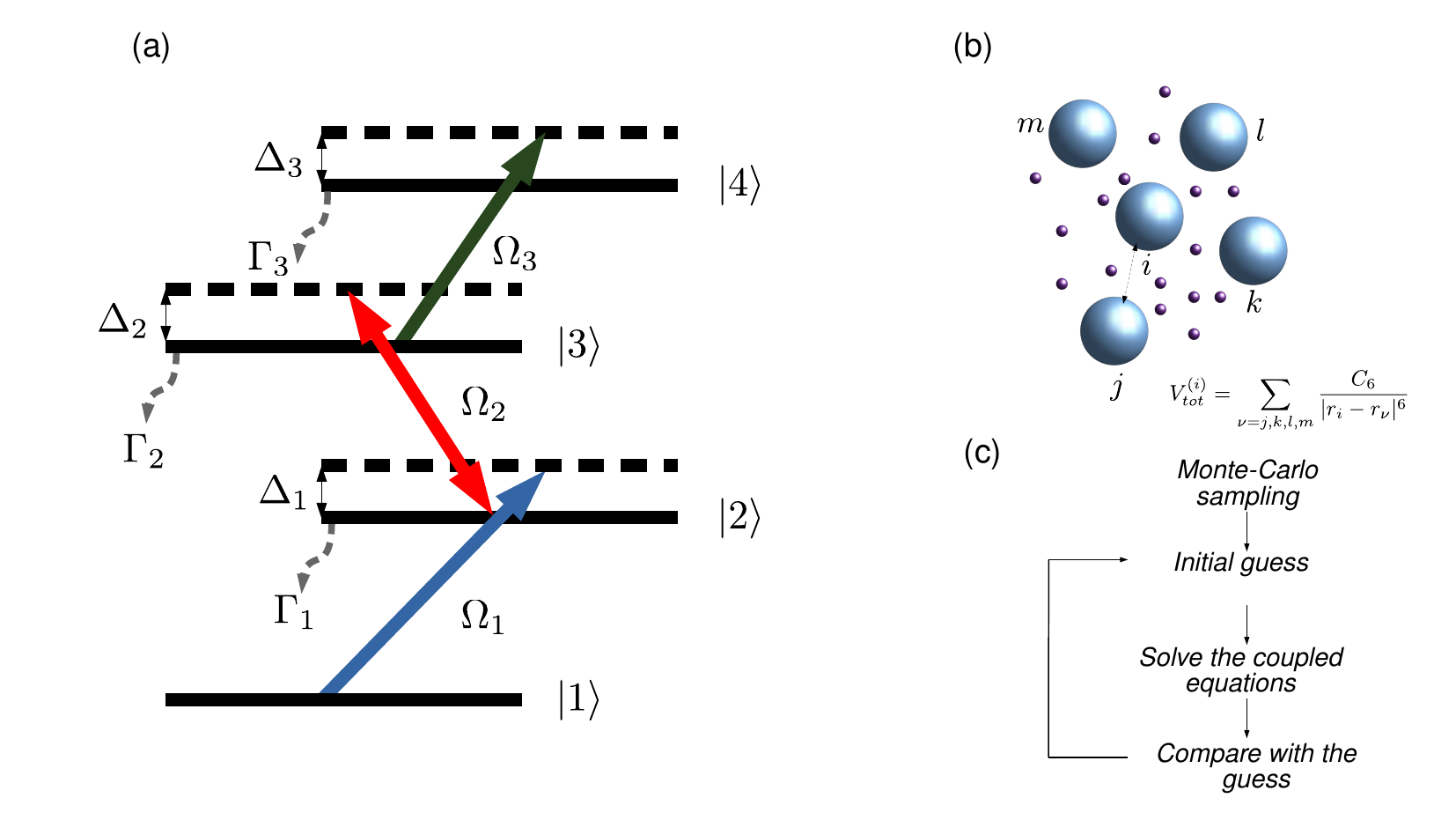}
	\caption{\label{ladder}(a) Three-photon excitation scheme. Excited state $|4\rangle$ is considered as a Rydberg state.(b) Schematic representation of Rydberg-Rydberg interactions. Big blue spheres are considered as Rydberg atoms while small purple ones are other atoms. (c) Self-consistent mean-field algorithm.}
\end{figure}

Few years ago, it was proposed that these effects which are generally observed via two-photon excitation, can also be realized with three-photon excitation scheme \cite{Sibalic2016}. Some recent studies investigated the dynamic dressing via three-photon excitation \cite{gao19} and transient nonlinear response of four-level Rydberg-EIT \cite{guo20}. Moreover, an experiment conducted with a Rb vapour cell demonstrated Rydberg EIA/EIT cross-over with this scheme \cite{taich19}. These new systems may have important contributions to applications such as coupling with optical fibers and more efficient and longer storage of light without decoherence effects. However, a theoretical model that could explain the experiments completely has yet to be developed. A realistic modeling of these systems is needed to fully understand the physics behind the phenomenon. So far, several theoretical models have been proposed to investigate three-level Rydberg-EIT systems and successfully explained experimental findings \cite{sev11b, ates11,  schem10,first16, petros11, heeg12}. Addition of one more atomic level into the system, significantly increases the dimension of Hilbert space and therefore it becomes very problematic to investigate four-level systems with realistic parameters in many-body picture with these methods.  To overcome this difficulty, we propose a self-consistent mean-field (MF) model to study four-level Rydberg atom systems under EIT and EIA conditions and, investigate the effects of strong inter-atomic interactions on quantum coherences and possible nonlinear optical effects. In this manuscript, we consider a collection of cold Cs and Rb atoms for Rydberg-EIT and Rydberg-EIA systems, respectively. Our simulations yield results that are consistent with the experiments which were performed with vapour cells and EIT-EIA cross-over is observed for Rb atoms by changing the system parameters. In other words, it is possible to change an absorptive medium to one which is tranparent. This kind of control on the system might be important for many nonlinear optics applications, since one can change the optical properties of the system as needed.

\section{System}
The system under consideration consists of atoms in three-photon ladder scheme in which atoms in the ground state are excited to a subsequent higher energy states via three distinct light fields as shown in figure \ref{ladder}(a). The ground state is labeled as $|1\rangle$ and the three following excited states are represented with $|2\rangle$, $|3\rangle$ and, $|4\rangle$ which corresponds to the Rydberg state. Excitation to higher energy states are accomplished via fields with respective Rabi frequencies, $\Omega_1$ (probe), $\Omega_2$ (dressing) and $\Omega_3$ and detunings, the difference between transition frequency and the frequency of the light fields, are shown with $\Delta_1$, $\Delta_2$ and $\Delta_3$. Spontaneous decay rates of the states are represented with $\Gamma_1$, $\Gamma_2$ and $\Gamma_3$, respectively and due to long lifetimes of Rydberg states $\Gamma_3$ is ignored in the simulations. We consider a cold Rydberg gas in frozen gas limit, in which atoms do not move within relevant timescales. Rydberg gases
are generally loaded into magneto-optical or dipole traps after cooling, however traps are turned off before the excitation scheme to avoid Stark shift. Therefore, we do not consider any effects due to trapping.

The equations of motion are obtained using the master equation

\begin{equation}\label{NbodyMastereqn}
	\dot{\rho}^{(N)} = -\frac{i}{\hbar}[\hat{H},\rho^{(N)}] + \mathcal{L}[\rho^{(N)}]\;\; ,
\end{equation}
where $\rho^{(N)}$ is the $N-$particle density matrix and the decay mechanisms are included in the Lindblad terms $\mathcal{L}[\rho^{(N)}]$. The total Hamiltonian consists of atomic and interaction parts, $\hat{H}=\hat{H}_A+\hat{V}$. Atomic Hamiltonian within rotating wave approximation is given as 

\begin{eqnarray}\label{totHamiltonian}
\hat{H}_A= \hbar \sum_{i=1}^{N} \left[ \frac{\Omega_1}{2} \hat{\sigma}_{12}^i + \frac{\Omega_2}{2} \hat{\sigma}_{23}^i  + \frac{\Omega_3}{2} \hat{\sigma}_{34}^i  -\Delta_1 \hat{\sigma}_{22}^i  \right.\nonumber\\ 
\left.  - (\Delta_1+\Delta_2)\hat{\sigma}_{33}^i -(\Delta_1+\Delta_2+\Delta_3)\hat{\sigma}_{44}^i  +h.c.\right]
\;\; , 
\end{eqnarray}
where $\hat{\sigma}_{\alpha\beta}^i=|\alpha\rangle\langle \beta|$ is the atomic transition operator for the $i^{th}$ atom. The interaction part, $\hat{V}$ is given as

\begin{equation}\label{vdwInteraction}
\hat{V} = \sum_{i\neq j} V_{ij} \hat{\sigma}_{44}^i \hat{\sigma}_{44}^j
\;\; .
\end{equation}
We consider inter-atomic interactions in the form of van der Waals (vdW) interactions, $ V_{ij}=-\frac{C_6}{{|\mathbf{r}_i - \mathbf{r}_j|}^6}$ where $C_6$ is the vdW coupling. One can obtain the equations of motion for reduced density matrices straightforwardly,
\begin{small}
\begin{numparts}
\begin{eqnarray}
\frac{d}{dt}\rho_{11}^{(i)}&=\frac{i}{2}\Omega_1^{(i)}\left(\rho_{12}^{(i)}-\rho_{21}^{(i)}\right)+\Gamma_{1}\rho_{22}^{(i)},\\	
\frac{d}{dt}\rho_{22}^{(i)} &=- \frac{i}{2}\Omega_1^{(i)}\left(\rho_{12}^{(i)}-\rho_{21}^{(i)}\right)+\frac{i}{2}\Omega_2^{(i)}\left(\rho_{23}^{(i)}-\rho_{32}^{(i)}\right)\\\nonumber
&+\Gamma_{2}\rho_{33}^{(i)}-\Gamma_{1}\rho_{22}^{(i)},\\		
\frac{d}{dt}\rho_{33}^{(i)} &=-\frac{i}{2}\Omega_2^{(i)}\left(\rho_{23}^{(i)}-\rho_{32}^{(i)}\right)+\frac{i}{2}\Omega_3^{(i)}\left(\rho_{34}^{(i)}-\rho_{43}^{(i)}\right) \\	\nonumber
&+\Gamma_{3}\rho_{44}^{(i)} -\Gamma_{2}\rho_{33}^{(i)},\\
\frac{d}{dt}\rho_{44}^{(i)} &= -\frac{i}{2}\Omega_3^{(i)}\left(\rho_{34}^{(i)}-\rho_{43}^{(i)}\right)-\Gamma_{3}\rho_{44}^{(i)},  \\	
\frac{d}{dt}\rho_{12}^{(i)} &= \frac{i}{2}\Omega_1^{(i)}\left(\rho_{11}^{(i)}-\rho_{22}^{(i)}\right)+\frac{i}{2}\Omega_2^{(i)}\rho_{13}^{(i)} - i\Delta_1^{(i)}\rho_{12}^{(i)}  \\	\nonumber
&- \frac{\Gamma_{1}}{2}\rho_{12}^{(i)}, \\
\frac{d}{dt}\rho_{13}^{(i)} &= \frac{i}{2}\Omega_3^{(i)}\rho_{14}^{(i)}+\frac{i}{2}\Omega_2^{(i)}\rho_{12}^{(i)}-\frac{i}{2}\Omega_1^{(i)}\rho_{23}^{(i)} \\\nonumber
&-i(\Delta_1^{(i)}+\Delta_2^{(i)})\rho_{13}^{(i)}-\frac{\Gamma_{2}}{2}\rho_{13}^{(i)}, \\	
\frac{d}{dt}\rho_{14}^{(i)} &= \frac{i}{2}\Omega_3^{(i)}\rho_{13}^{(i)}-\frac{i}{2}\Omega_1^{(i)}\rho_{24}^{(i)}-i(\Delta_1^{(i)}+\Delta_2^{(i)}+\Delta_3^{(i)})\rho_{14}^{(i)} \label{coh1}\\\nonumber
&-\frac{\Gamma_{3}}{2}\rho_{14}^{(i)}+i\sum\limits_{i\neq j} V_{ij}\rho_{14,44}^{(i,j)}, \\	
\frac{d}{dt}\rho_{23}^{(i)} &= \frac{i}{2}\Omega_2^{(i)}\left(\rho_{22}^{(i)}-\rho_{33}^{(i)}\right) -\frac{i}{2}\Omega_1^{(i)}\rho_{13}^{(i)}+\frac{\Omega_3^{(i)}}{2}\rho_{24}^{(i)} \\\nonumber
&-i\Delta_2^{(i)}\rho_{23}^{(i)}-\frac{(\Gamma_{1}+\Gamma_{2})}{2}\rho_{23}^{(i)},\\	
\frac{d}{dt}\rho_{24}^{(i)} &= \frac{i}{2}\Omega_3^{(i)}\rho_{23}^{(i)}-\frac{i}{2}\Omega_2^{(i)}\rho_{34}^{(i)}-\frac{i}{2}\Omega_1^{(i)}\rho_{14}^{(i)}\\\nonumber
&-i(\Delta_2^{(i)}+\Delta_3^{(i)})\rho_{24}^{(i)} -\frac{(\Gamma_{1}+\Gamma_{3})}{2}\rho_{24}^{(i)}+i\sum\limits_{i\neq j} V_{ij}\rho_{24,44}^{(i,j)}, \label{coh2} \\	
\frac{d}{dt}\rho_{34}^{(i)} &= \frac{i}{2}\Omega_3^{(i)}\left(\rho_{33}^{(i)}-\rho_{44}^{(i)}\right) -\frac{i}{2}\Omega_2^{(i)}\rho_{24}^{(i)}-i\Delta_3^{(i)}\rho_{34}^{(i)} \label{coh3}\\\nonumber
&-\frac{(\Gamma_{2}+\Gamma_{3})}{2}\rho_{34}^{(i)}+i\sum\limits_{i\neq j} V_{ij}\rho_{34,44}^{(i,j)}	.
\end{eqnarray}
\end{numparts}
\end{small}
To be able to solve these coupled differential equations, one needs two-body density matrices $\rho_{\alpha 4,44}^{(i,j)}$. Evaluating two-body density matrices requires knowledge on three-body density matrices and they, in turn involve four-body density matrices and it goes on in the same hierarchy. Therefore, it is  not possible to solve these equations exactly for big systems and the problem requires a different approach. There have been many theoretical proposals to deal with this problem for three-level scheme \cite{ates11, sev11b, schem10,first16, petros11, heeg12}, and in this manuscript, we propose a self-consistent MF approach to overcome the difficulties arising from computational power and numerical instability problems due to increase in the size of Hilbert space.
 
\subsection{Self-consistent MF approach}
MF approximation, in which it is possible to reduce two-body terms into products of single-body density matrices that we have knowledge of has been used widely. However, it is shown that MF approximation underestimates the correlations between the Rydberg atoms and therefore does not yield accurate results for all parameter regimes \cite{schem10}. MF approach would also fail for very small atomic densities where fluctuations in the fields experienced by atoms become pronounced. Thus, we propose a MF approach that includes self-consistency to overcome this problem and develop an algorithm that involves the correlations. Firstly, we reduce two-body density matrices to the product of single-body density matrices by MF approximation as $\rho_{\alpha 4,44}^{(i,j)} =\rho_{\alpha 4}^{(i)}\rho_{44}^{(j)}$. Only off-diagonal elements of the density matrix, i.e. coherences, (\ref{coh1}-\ref{coh3}) include interaction terms, so modified equations become 
\begin{small}
\begin{numparts}
\begin{eqnarray}
\frac{d}{dt}\rho_{14}^{(i)}&=\frac{i}{2}\Omega_3\rho_{13}^{(i)}-\frac{i}{2}\Omega_1\rho_{24}^{(i)}-i(\Delta_1+\Delta_2+\Delta_3)\rho_{14}^{(i)}\nonumber\\
& -\frac{\Gamma_{3}}{2}\rho_{14}^{(i)}+i\sum\limits_{i\neq j} V_{ij}\rho_{14}^{(i)}\rho_{44}^{(j)}, \label{MFA1}\\
\frac{d}{dt}\rho_{24}^{(i)}&=\frac{i}{2}\Omega_3\rho_{23}^{(i)}-\frac{i}{2}\Omega_2\rho_{34}^{(i)}-\frac{i}{2}\Omega_1\rho_{14}^{(i)}-i(\Delta_2+\Delta_3)\rho_{24}^{(i)}\nonumber \\
 & -\frac{(\Gamma_{1}+\Gamma_{3})}{2}\rho_{24}^{(i)}+i \sum\limits_{i\neq j}V_{ij}\rho_{24}^{(i)}\rho_{44}^{(j)},\label{MFA2}\\
\frac{d}{dt}\rho_{34}^{(i)}&=\frac{i}{2}\Omega_3\left(\rho_{33}^{(i)}-\rho_{44}^{(i)}\right) -\frac{i}{2}\Omega_2\rho_{24}^{(i)}-i\Delta_3\rho_{34}^{(i)} \nonumber \\ 
&-\frac{(\Gamma_{2}+\Gamma_{3})}{2}\rho_{34}^{(i)}+i\sum\limits_{i\neq j} V_{ij}\rho_{34}^{(i)}\rho_{44}^{(j)}\label{MFA3}.
\end{eqnarray}
\end{numparts}
\end{small}
Since all interaction terms in \ref{MFA1}-\ref{MFA3} have Rydberg state population, $\rho_{44}$ as a common term, one can write it for $i^{th}$ atom as
 \begin{equation}
 i\sum\limits_{i\neq j} V_{ij}\rho_{\alpha 4}^{(i)}\rho_{44}^{(j)} = -i \rho_{\alpha 4}^{(i)} \delta^{(i)},
 \end{equation}
where $\delta^{(i)} = \sum\limits_{j\neq i}V_{ij} \rho^{(j)}_{44}$ represents the sum of interactions with all other Rydberg atoms. Figure \ref{ladder}(b) shows the schematic representation of how total interaction for the $i^{th}$ Rydberg atom is calculated.
 Following that, we start with an initial guess for $\rho_{44}$, e.g. population for the non-interacting case, and calculate pairwise interactions for randomly distributed atoms using Monte-Carlo sampling for atomic positions. By introducing interactions to the system, we solve the system of linearly coupled equations for $\rho_{44}$ self-consistently and obtain the steady-state solutions. The self-consistent MF algorithm is shown in figure \ref{ladder}(c).

\section{Results and Discussion} 

\subsection{Rydberg-EIT}

For the system composed of cold Cs atoms, we choose the state configuration as: $6S_{1/2} \rightarrow 6P_{3/2} \rightarrow 7S_{1/2} \rightarrow nP$ and the parameters as: $(\Omega_1,\Omega_2,\Omega_3)/(2\pi)=(0.1,8,1)$ MHz,  $(\Gamma_{1},\Gamma_{2},\Gamma_{3})/(2\pi)=(5.39, 3.31, 0)$ MHz and, $(\Delta_2, \Delta_3)/(2\pi)=(0,-4)$ MHz in accordance with the experiment \cite{Sibalic2016} . Simulations are completed for $50$ atoms and averaged over 1000 realizations of atomic positions in a given atomic density $\varrho=10^9$ cm$^{-3}$ with periodic boundary conditions. Since the interaction strength between the Rydberg atoms scales with principal quantum number as $C_6\propto n^{11}$, figures are plotted for  $n=50,60,70,80,90,100$ to visualize the interaction effects. 

\begin{figure}[!htbp]
    \centering
	\includegraphics[width= 10 cm]{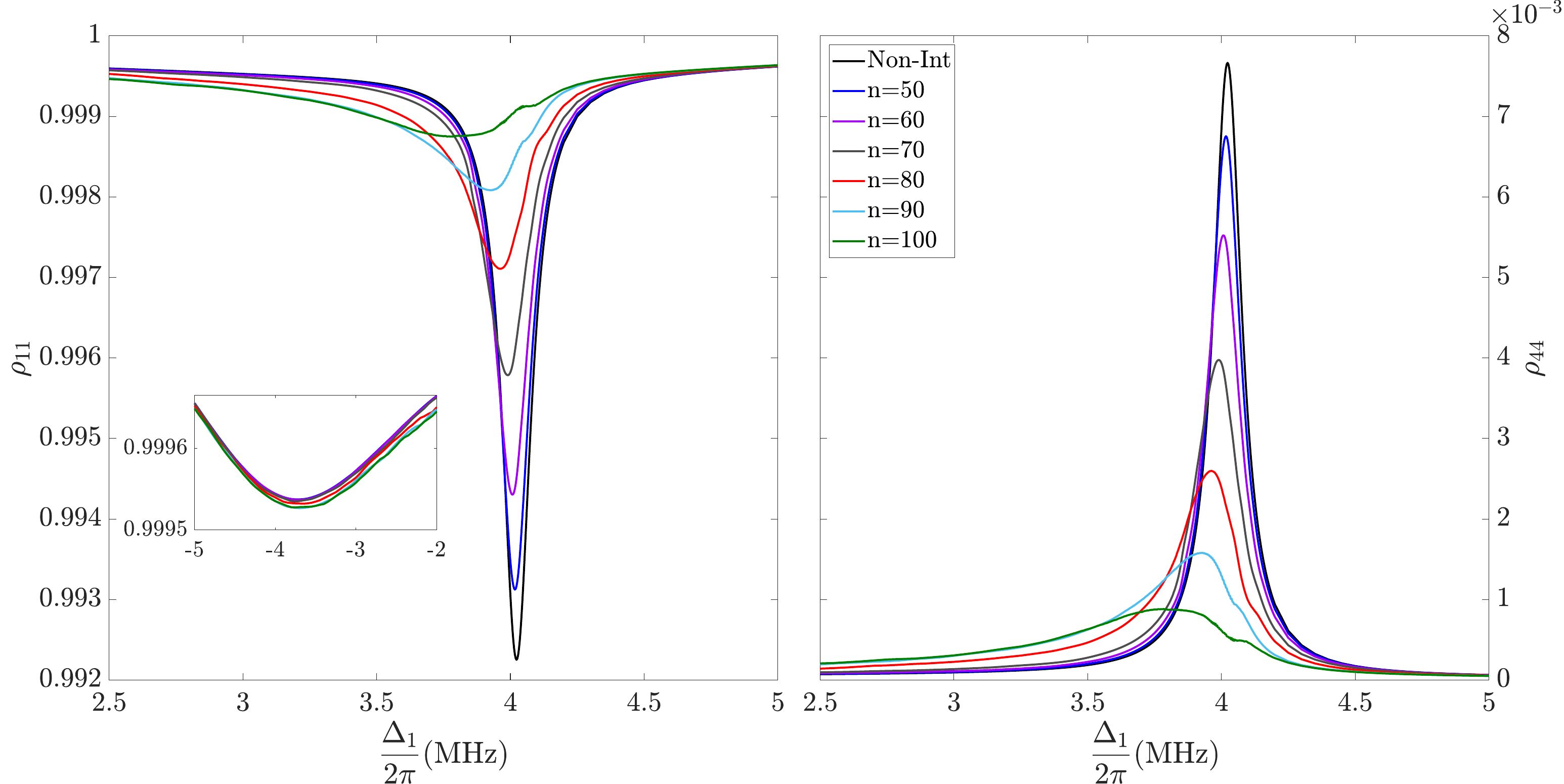}
	\caption{Ground-state (left) and Rydberg state (right) populations as a function of probe detuning for different interaction strengths. Inset zooms around $\Delta_1 /(2\pi)=- 4$ MHz.}\label{g-ryd_pop}
\end{figure}

Figure \ref{g-ryd_pop} shows the ground state (left) and Rydberg state (right) populations for different interaction strengths, scanned across probe field detuning $\Delta_1$. Inset shows the region zoomed around $\Delta_1/(2\pi)=-4$ MHz. Since the probe field is weak, most of the population in the system is at ground state. At three-photon resonance ($\Delta_1+\Delta_2+\Delta_3=0$), ground state population has the lowest value because excitation to Rydberg state is most effectively achieved with this condition. But, as the principal quantum number $n$ increases, probability of exciting the atoms decreases, effectively trapping the atoms at the ground state. As the interaction increases, Rydberg blockade radius gets larger and more atoms are left inside the blockaded region, therefore reducing the number of atoms in a Rydberg state at the same time. This excitation suppression, traps atoms in the ground state, crippling the dark state formation, thus EIT and dispersive feature lose their prominence as it can be seen from the following figures.

\begin{figure}[!htbp]
    \centering
	\includegraphics[width= 10 cm]{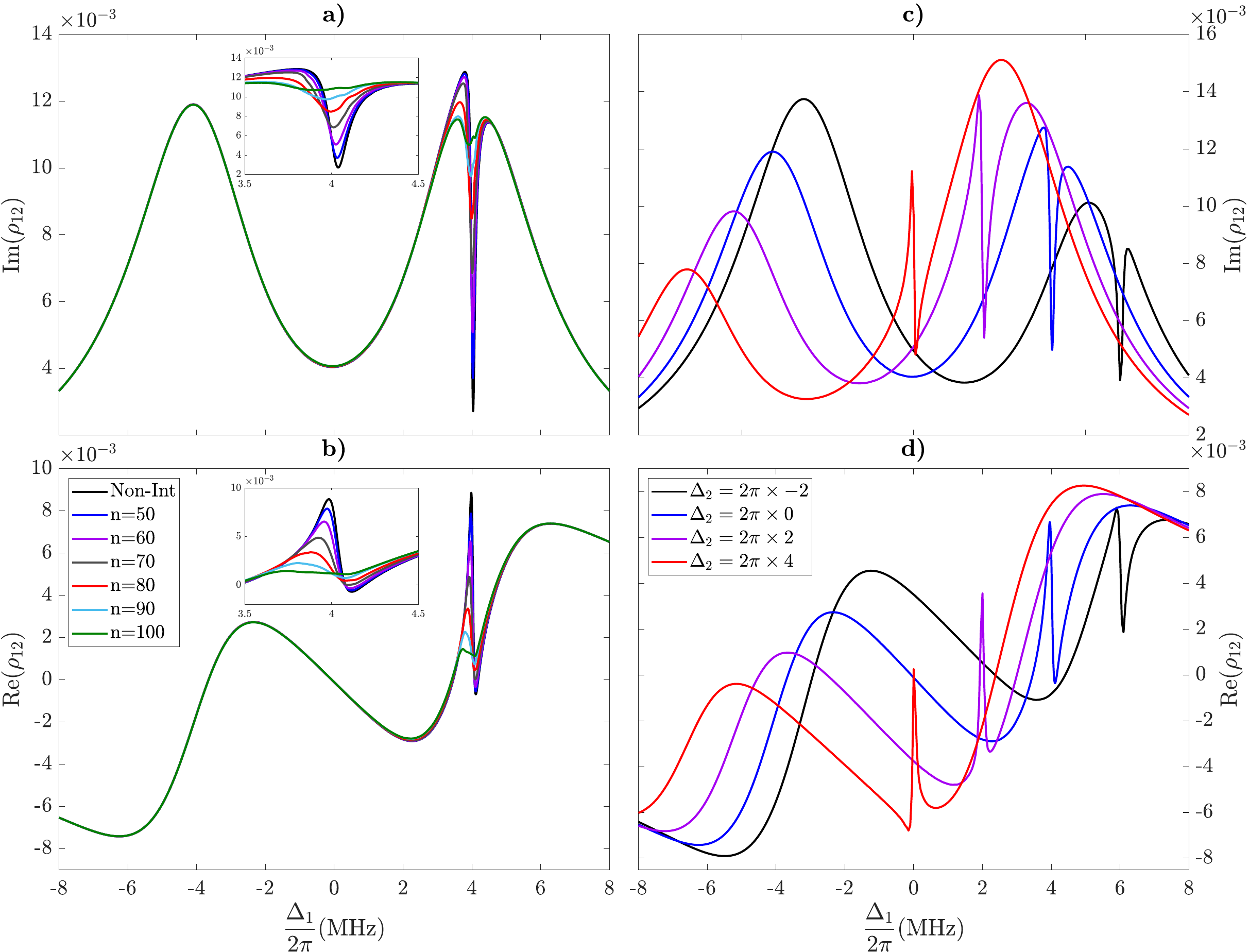}
	\caption{Imaginary and real parts of the probe coherence for different interaction strengths (a, b) and different coupling detunings (c,d). Insets in (a) and (b) zoom around the three-photon resonance. }\label{coh}
\end{figure}

Figure \ref{coh}(a) shows the occurence of transparency due to the strong dressing around three-photon resonance. Two absorption peaks at $\Delta_1/( 2 \pi)=-4$ and $4$ MHz correspond to the absorption of the dressed states. But due to the dark state formation at three-photon resonance, absorption minima is observed instead of a peak. Figure \ref{coh}(b) shows the dispersive feature introduced by EIT mechanism at three-photon resonance. In the non-interacting case, steep dispersion curve is observed, but as the van der Waals interaction gets stronger, slope of the dispersion curve gets smoother and at $n=100$ as can be seen from the inset, steep curvature is completely lost.

In figures \ref{coh}(c) and (d) imaginary and real parts of probe coherence for different dressing field detunings with respect to probe field detuning are shown, respectively while van der Waals interaction is kept constant with $n=60$. We expected to observe EIT behavior at three-photon resonance and it can be seen that EIT and steep change in refractive index, shift towards three-photon resonance with changing $\Delta_2$. It is clear that three-photon resonance is required for formation of dark-state and transparency window to occur. 

\begin{figure}[!htbp]
    \centering
	\includegraphics[width= 10 cm]{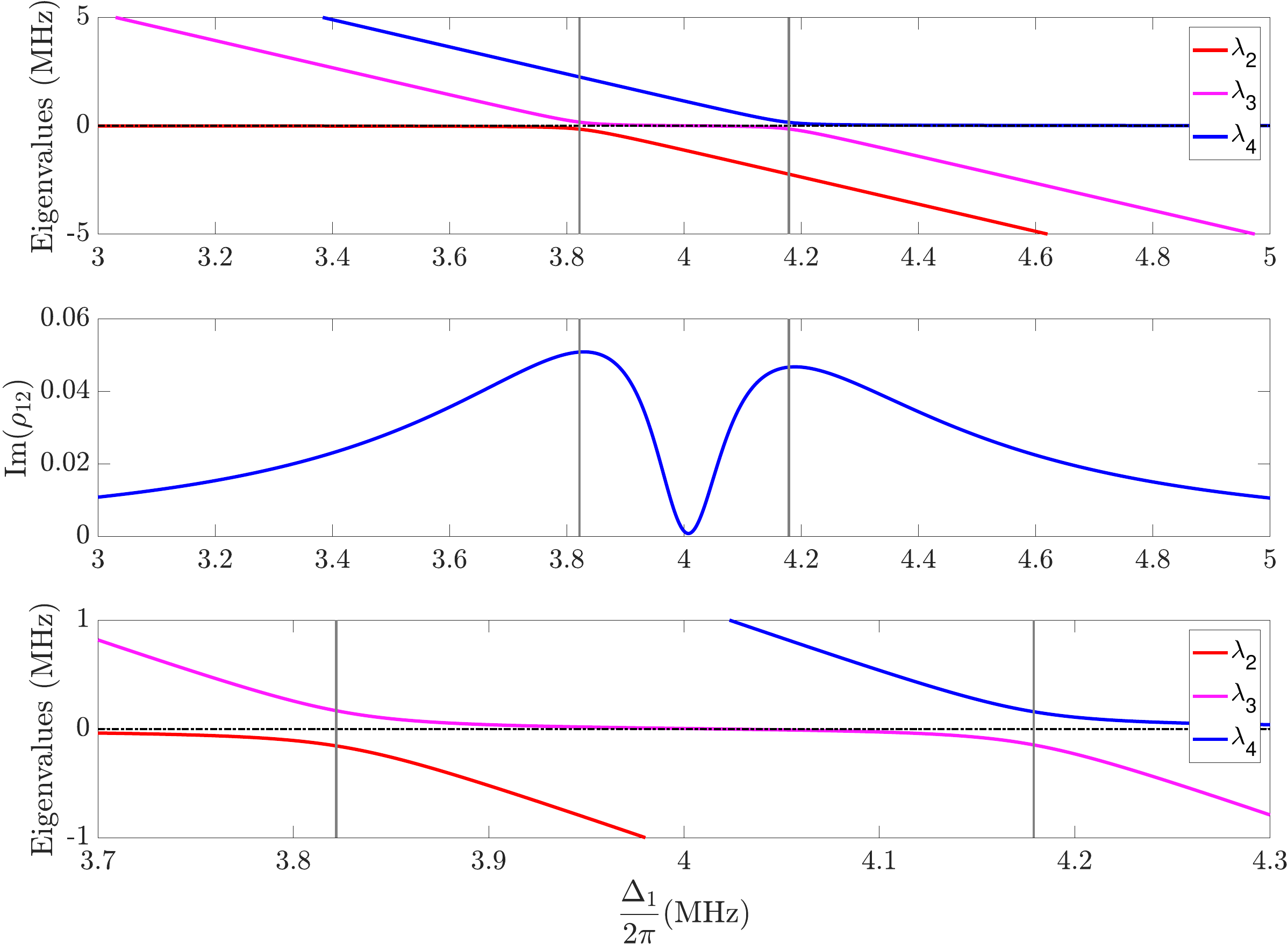}
	\caption{Hamiltonian eigenvalues $\lambda_2 , \lambda_3 , \lambda_4$ (top, bottom) and the imaginary part of the probe coherence (middle). $\lambda_1$ is not shown here since it is zero. Vertical grey lines are the maximum absorption points around the three-photon resonance. Bottom panel zooms in around three-photon resonance.}\label{eigen}
\end{figure}

As it can be seen from figure  \ref{eigen}, strong dressing field $\Omega_2$ together with a weak probe field $\Omega_1$ in ladder configuration results in avoided crossings and dark-state formation at three-photon resonance which permits a transparency window to be opened. Figure \ref{eigen} shows the eigenvalues of the Hamiltonian (top, bottom) without interactions as a function of probe detuning together with the imaginary part of the probe coherence  (middle). There are avoided crossings at the points shown with grey vertical lines where absorption is maximum. System evolves into a dark state between these points and EIT window appears around three-photon resonance in accordance with the reference\cite{gao19}. The inclusion of atomic interactions makes it clear that EIT is prone to be disturbed. It might be helpful to study the energy eigenvalues and probe coherence as a function of probe detuning for a deeper understanding of the underlying mechanism.  

\subsection{Rydberg-EIA}
To investigate Rydberg-EIA system, we now consider a cold Rb gas cloud and treat the system with self-consistent MF approach to obtain steady-state solutions for different interaction strengths. The excitation scheme under examination is 
$5S_{1/2}\rightarrow 5P_{3/2}\rightarrow 5D_{5/2}\rightarrow nF$ and the parameters used in the simulations are $(\Omega_1,\Omega_2,\Omega_3)/( 2\pi)= (10,25,18)$ MHz,  $(\Gamma_1,\Gamma_2,\Gamma_3)/( 2\pi)= (6,0.66,0)$ MHz which are consistent with the experiment \cite{taich19}. In the experiment, Rb vapour cell was used and the Rydberg state is coupled with another one with a microwave (MW) field for the non-interacting system. However here, we consider a cold interacting gas instead, to investigate the interaction effects.  Simulations are completed for the detunings $\Delta_1 = 0$, $\Delta_2/(2\pi)= (-20, 0,20)$ MHz and atomic density $\varrho=10^9$ cm$^{-3}$.

\begin{figure}[!htbp]
    \centering
	\includegraphics[width= 10 cm]{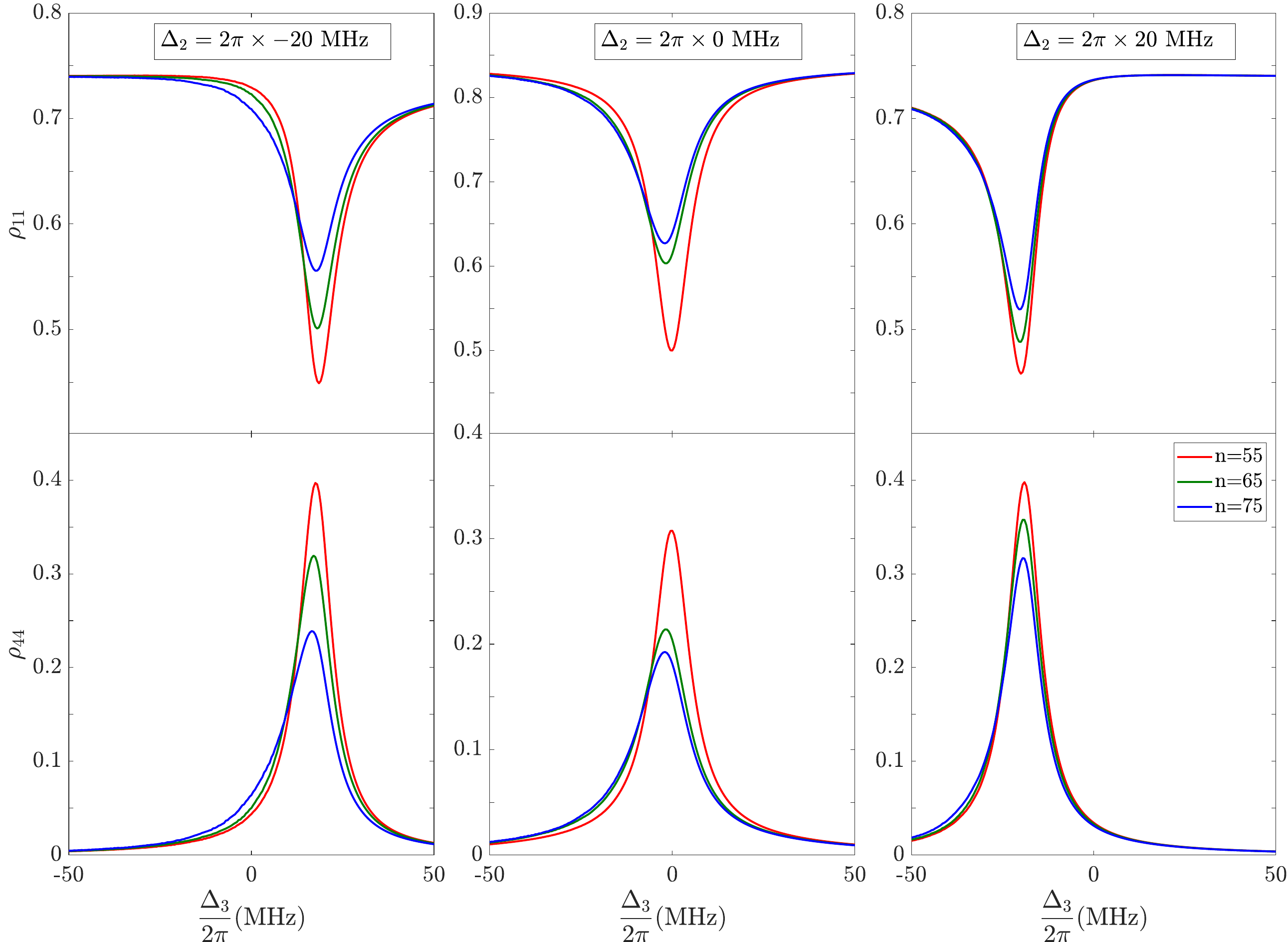}
	\caption{Ground-state (top) and Rydberg state (bottom) populations as a function of dressing detuning for different interaction strengths. Each column is for different coupling detuning; $\Delta_2/(2\pi)= (-20, 0,20)$ MHz (left, center, right).}\label{eiapops}
\end{figure}

Figure \ref{eiapops} shows the ground state (top) and Rydberg state (bottom) populations for different interaction strengths with  $\Delta_2/(2\pi)= (-20, 0,20)$ MHz (left, center, right). Since we consider a weak probe field, the population mostly stays on the ground state and Rydberg state excitation happens at the three-photon resonance as in the case of Rydberg-EIT system. Interaction reduces the Rydberg population as expected and the population is trapped in the ground-state as the interaction increases. Blockade effect \cite{gaet09,urb09,comp10} can be seen from the figure very clearly for all $\Delta_2$ values as well as importance of the three-photon resonance. 

\begin{figure}[!htbp]
    \centering
	\includegraphics[width= 10 cm]{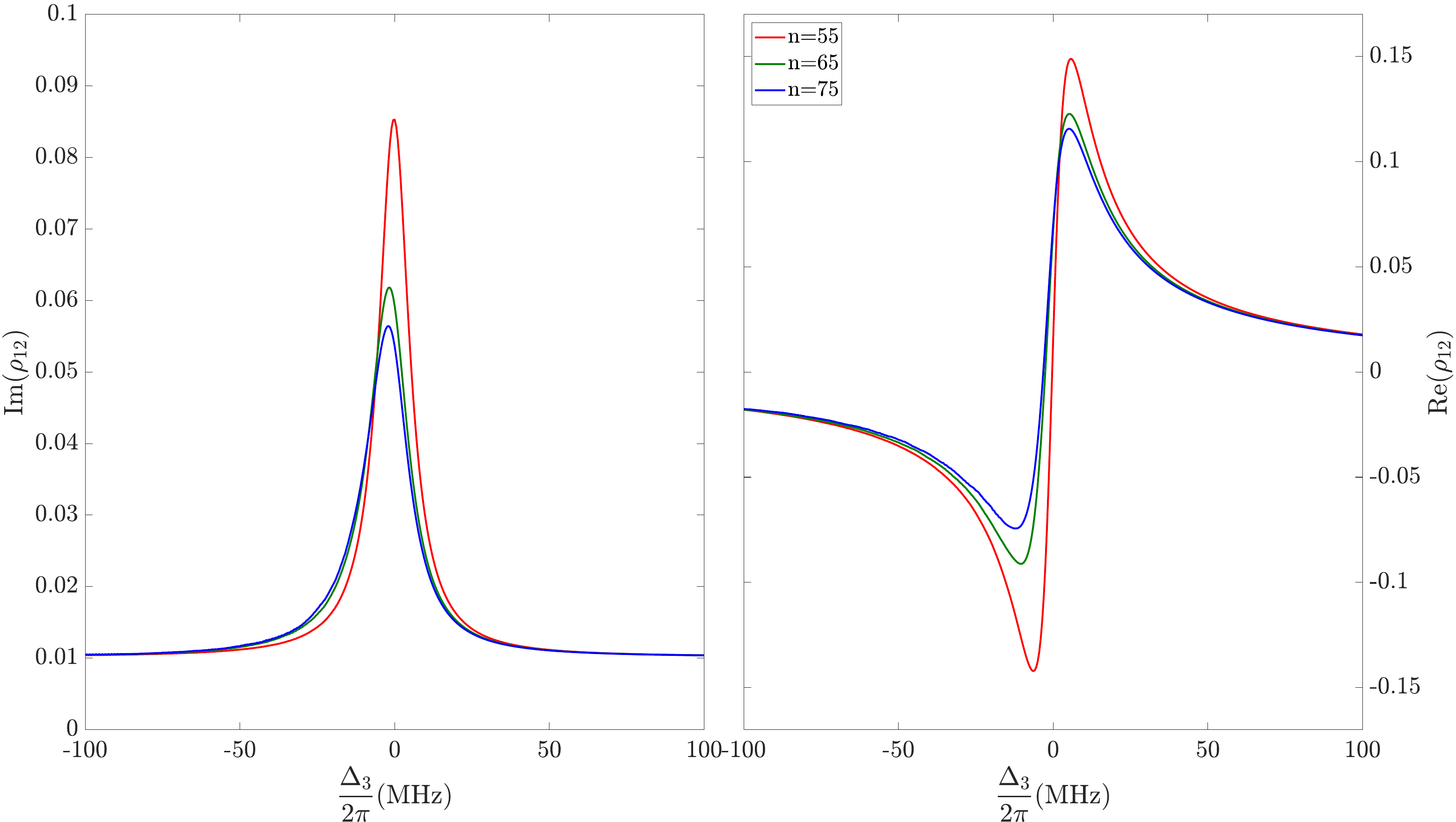}
	\caption{Imaginary (left) and real (right) parts of the probe coherence as a function of dressing detuning for different interaction strengths.}\label{eiacoh_n}
\end{figure}

Figure \ref{eiacoh_n} shows the imaginary (left) and real (right) parts of the probe coherence as a function of $\Delta_3$ for different interaction strengths. The absorption peak is observed at  $\Delta_3= 2\pi\times 0$ MHz and as the interaction and therefore the blockade radius increase, there is a reduction in absorption and system is expected to behave like an effective-three level system for high principal quantum numbers. The real part determines the dispersive behavior and the slope of it plays an important role for slow-light applications. Interaction strongly affects the slope, so it might be possible to change the refractive index significantly.  

\subsection{EIT-EIA cross-over}

\begin{figure}[!htbp]
    \centering
	\includegraphics[width= 10 cm]{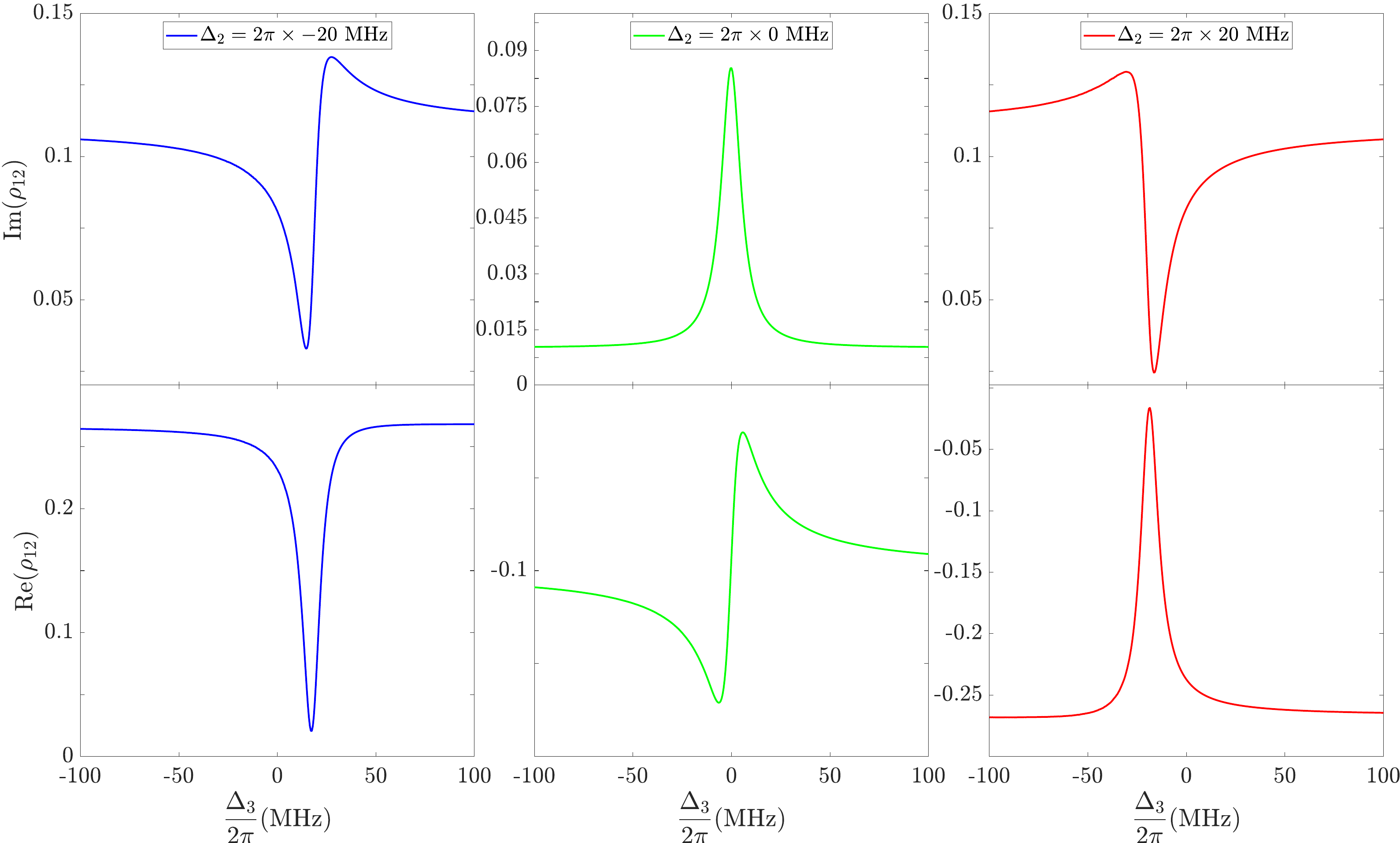}
	\caption{EIT-EIA crossing for different coupling detunings; $\Delta_2/(2\pi)=(-20,0,20)$ MHz (left, center, right). Top panels show the imaginary part of the coherence whereas bottom ones show the real part.}\label{eit-eia}
\end{figure}

Our simulations also demonstrate that it is possible to have EIA-EIT cross-over by changing the dressing detuning as it was observed in experiment \cite{taich19}.  Figure \ref{eit-eia} shows the imaginary (top) and real (bottom) parts of the probe coherence for different coupling detunings for $n=55$ state. In the case of $\Delta_2/(2\pi)= 0$ (center) there is an EIA signal whereas for the cases  $\Delta_2/(2\pi)=(-20,20)$ MHz (left,right) there appears an EIT window instead. It is obvious that only by changing the detuning, one can observe EIT-EIA cross-over and therefore it is quite easy to manipulate the system. It is clearly seen that if all fields are coupled resonantly, i.e. when all detunings are zero, system exhibits EIA whereas off-resonant coupling of probe and coupling fields, results in EIT. This kind of control on the system might provide a very useful platform for many applications, since it is possible to change the optical properties of the system as needed.  The refractive index of the system exhibits a drastic change as it can be seen from the figure. The three-photon resonance is still significantly important and we consider non-interacting case to avoid screening effects of it on EIT-EIA cross-over. In the case of EIT, real part of the probe coherence shows a very different dispersive effect compared to the Cs case (figure \ref{coh}(b)). It might be interesting to study this behavior further in the future.  

\begin{figure}[!htbp]
    \centering
	\includegraphics[width= 10 cm]{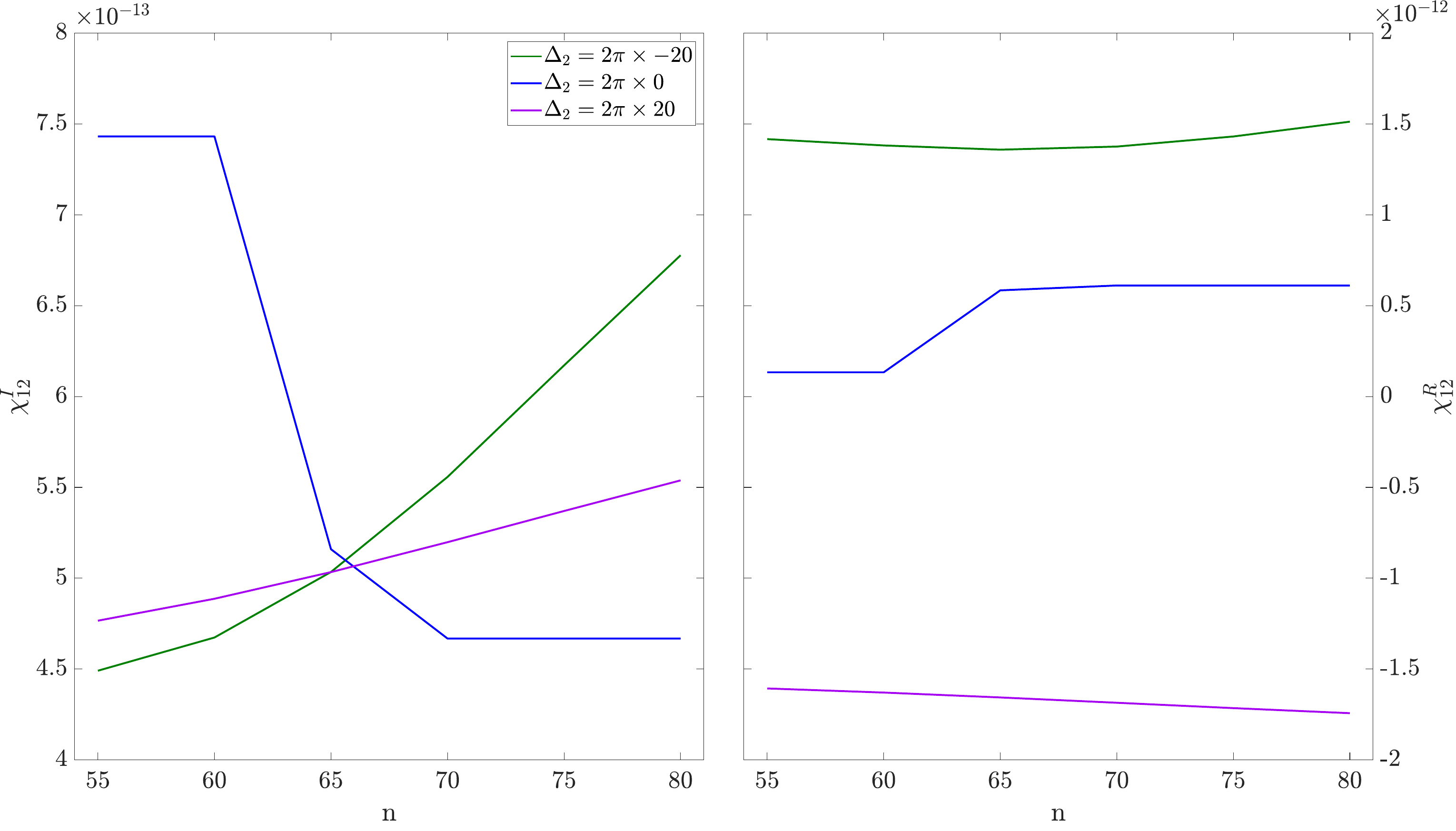}
	\caption{Imaginary (left) and real (right) parts of the susceptibility as a function of n for EIT ($\Delta_2/(2\pi)=-20,20$) and EIA ($\Delta_2/( 2\pi)=0$) regimes.}\label{suscep}
\end{figure}

Finally, the susceptibility, $\chi_{12}=\chi_{12}^R+\chi_{12}^I= 2\varrho^2\rho_{12}/(\hbar\epsilon_0\Omega_1)$ for atomic density $\varrho=10^9$ cm$^{-3}$,  is shown in figure \ref{suscep} as a function of interaction strengths for different regimes which also shows that optical properties of these systems can be easily manipulated by changing the parameters. As the coupling detuning varies, optical behavior changes completely and there is a high loss with a steep slope on the real part at three-photon resonance in the case of EIA. However, for both EIT cases, it is possible to reduce the group velocity without high loss. The difference between EIA and EIT is more clearly seen here and the interaction has drastic effects on the susceptibility as expected. As the interaction increases, loss is reduced and saturated in the case of EIA, whereas in the EIT cases, loss is increased with it. Saturation is observed more clearly on the real part (right). More in depth simulations are required to investigate nonlinear effects but even with these basic susceptibility calculations, it is obvious that such systems can easily be manipulated on demand.

\section{Conclusion}
We investigated the interaction effects on three-photon Rydberg-EIT and Rydberg-EIA phenomena by using the self-consistent MF approach. Many-body simulation results agree with experimental findings quantitatively and,  show that it is easy to manipulate these systems with system parameters such as field detunings and principal quantum number for different optical modification purposes. EIT-EIA cross-over that was observed experimentally before \cite{taich19} is also studied for Rb systems. Four-level
excitation scheme would provide a potential platform for applications
such as THz sensing and imaging and, Rydberg spectroscopy because of the control on the optical response of atomic media. Three-photon excitation scheme might be experimentally more
feasible when compared to systems using two-photon excitation, due to the
availability of infrared diode lasers.  It is also possible to achieve higher Rabi frequencies using low
laser power. Although subject of Rydberg-EIT is studied for more than a decade, comprehensive work on four-level
Rydberg-EIT systems are rare. Thus, our method provides a practical tool to deeply understand these quantum coherence phenomena and further investigation on susceptibilities for nonlinear effects might provide an insight for applications.

\ack
S. Sevin\c{c}li and Y. Oyun acknowledge support from the Scientific and Technological Research Council of Turkey (TUBITAK) under  Grant  No.  117F372. 


\textbf{}\bibliography{biblio}

\end{document}